# Power Allocation and Measurement Matrix Design for Block CS-Based Distributed MIMO Radars

Azra Abtahi, Mahmoud Modarres-Hashemi, Farokh Marvasti, and Foroogh S. Tabataba

*Abstract*— **Multiple-input multiple-output (MIMO) radars offer higher resolution, better target detection, and more accurate target parameter estimation. Due to the sparsity of the targets in space-velocity domain, we can exploit Compressive Sensing (CS) to improve the performance of MIMO radars when the sampling rate is much less than the Nyquist rate. In distributed MIMO radars, block CS methods can be used instead of classical CS ones for more performance improvement, because the received signal in this group of MIMO radars is a block sparse signal in a basis. In this paper, two new methods are proposed to improve the performance of the block CS-based distributed MIMO radars. The first one is a new method for optimal energy allocation to the transmitters, and the other one is a new method for optimal design of the measurement matrix. These methods are based on minimizing an upper bound of the sum of the block-coherences of the sensing matrix blocks. Simulation results show an increase in the accuracy of multiple targets parameters estimation for both proposed methods.**

## I. INTRODUCTION

MULIPLE-INPUT multiple-output (MIMO) radar [1],[2] is a radar that uses multiple antennas to simultaneously transmit diverse waveforms and multiple antennas to receive the reflected signal. The received signals are sent to a common processing center that is called fusion center. The difference between a MIMO radar and a phased-array radar is that the MIMO radar can transmit multiple waveforms from its transmitters, while in a phased array radar various shifts of the same signal are transmitted. There are two kinds of MIMO radars: distributed MIMO radars and co-located MIMO radars. In co-located type [3],[4], transmitters and receivers are located close to each other relative to their distance to the target; thus all transmitter-receiver pairs view the target from the same angle. In co-located MIMO radars, the phase differences induced by transmitters and receivers can be used to form a long virtual array with the number of elements equal to the product of the number of transmitters and receivers; therefore, they can achieve superior Direction of Arrival (DOA) resolution [3]. In distributed MIMO radars [5]-[7], the transmitters are located far apart from each other relative to their distance to the target. In this type of MIMO radars, the target is viewed from different angles. Thus, if the received signal from a particular transmitter and receiver is weak, it can be compensated by the received signals from other transmitter-receiver pairs. This type of MIMO radars is shown to offer superior target detection, more accurate target parameter estimation, and higher resolution [1],[5]-[7].

If the sampling rate in MIMO radars is reduced, the cost of receivers can be reduced, and because of the existence of the multiple receivers, this reduction is very significant. Compressive Sensing (CS) methods make this reduction possible. Using this signal processing method, we can remove the need of the high rate A/D converters and send much less samples to the fusion center. Compressive sensing [8]-[14] is a new paradigm in signal processing that allows us to accurately reconstruct sparse or compressible signals from a number of samples which is much smaller than that is necessary according to the Shannon-Nyquist sampling theory. A vector $x$ is called $K$-sparse if it is a linear combination of only $K$ basis vectors. In other words, using the basis matrix $\Psi$ with basis vectors as columns, we can express $x$ as

$$x = \Psi s \qquad (1)$$

where $s$ is the weighting coefficient vector with length of $N$ in the basis $\Psi$, and if only $K$ elements of $s$ are nonzero, $x$ is called $K$-sparse [8-12]. Compressive sensing is more valuable when $K \ll N$. $x$ is compressible if it has just a few large coefficients and many small coefficients [11].

CS-based MIMO radars can be improved by different methods like: optimal design of measurement matrix [15],[16], and optimal design of transmitted waveforms [17],[18]-[20]. It is shown that these systems can estimate target parameters better than MIMO radars that are using some other estimation methods with higher sampling rates [15]-[17],[21],[22].

Let us consider $s$ as a concatenation of blocks with length $d$, i.e.,

A. Abtahi was with Department of Electrical and Computer Engineering, Isfahan University of Technology, Isfahan, Iran. She is now with Advanced Communication Research Institute (ACRI), EE Department, Sharif University of Technology, Tehran, Iran (e-mail: Azra_abtahi@ee.sharif.edu).
M. Modarres Hashemi and F. S. Tabataba are with Department of Electrical and Computer Engineering, Isfahan University of Technology, Isfahan, Iran (e-mails: Modarres@cc.iut.ac.ir; Fstabataba@cc.iut.ac.ir).
F. Marvasti is with Advanced Communication Research Institute (ACRI), EE Department, Sharif University of Technology, Tehran, Iran (e-mail: Fmarvasti@sharif.edu).

$$\mathbf{s} = [\underbrace{s_1, \ldots, s_d}_{S[1]}, \underbrace{s_{d+1}, \ldots, s_{2d}}_{S[2]}, \ldots, \underbrace{s_{N-d+1}, \ldots, s_N}_{S[N/d]}]^T \quad (2)$$

where $(.)^T$ denotes transpose of a matrix. If at most $K$ blocks of $\mathbf{s}$ have nonzero Euclidean norms, $\mathbf{x}$ is called block $K$-sparse in the basis $\mathbf{\Psi}$ [23]-[27]. For block-sparse signals it is better to use block CS methods instead of usual CS methods. References [19], [28], and [29] use block CS methods in distributed MIMO radars and show the advantages of these methods over using the classical CS ones. It should be noted that we can use block CS methods in this type of MIMO radar because the received signal in this system is block-sparse. Reference [19] proposed an adaptive energy allocation method for block CS-based distributed MIMO radars, too. However, so far, no non-adaptive method has been proposed for improving the performance of the block CS-based distributed MIMO radars in particular.

In block CS methods such as BOMP and BMP, a critical parameter named block coherence is required to be small enough for appropriate recovery [25]. This parameter is defined in section IV. For traditional CS methods, the coherence, the maximum correlation between the sensing matrix columns, is the determinative parameter. However, by reducing the coherence, the block coherence may also decrease. Reference [20] has used this idea to allocate energy to the transmitters in a block CS-based distributed MIMO radar. In this paper, we propose a superior transmitted energy allocation method to minimize an upper bound of the sum of the block-coherences of the sensing matrix blocks. We then, design a proper measurement matrix using the proposed upper bound as the cost function. We show that using these methods, the block CS-based distributed MIMO radar can be more accurate for target parameter estimation when the total transmitted energy is constant. The superiority of our proposed energy allocation method over the proposed method in [20] is also shown in the simulation results section.

The paper is organized as follows. In section II, we provide the received signal model of the CS-based distributed MIMO radar system. In section III, two important block recovery algorithms are presented. A new method of transmitted-energy allocation based on the sensing matrix block-coherence is introduced in section IV. In section V, we present a new method based on the optimization of the measurement matrix to improve the performance of the CS-based distributed MIMO radars. Section VI is allocated to simulation results, and finally, we make some concluding remarks in section VII.

## II. RECEIVED SIGNAL MODEL FOR CS-BASED DISTRIBUTED MIMO RADAR

Let us consider a distributed MIMO radar system consisting of $M_t$ transmitters and $N_r$ receivers. The $i^{th}$ transmitter and $l^{th}$ receiver are located at $\mathbf{t}_i = [t_{x_i}, t_{y_i}]$ and $\mathbf{r}_l = [r_{x_l}, r_{y_l}]$ on a Cartesian coordinate system, respectively. We transmit orthogonal waveforms of duration $T_P$ from different transmitters, and Pulse Repetition Interval (PRI) is $T$. $x_i(t)$ is a complex baseband waveform with energy equal to 1, and $p_i x_i(t) e^{j2\pi f_c t}$ is the waveform transmitted from the $i^{th}$ transmitter where $f_c$ is the carrier frequency. So, the transmitted energy from the $i^{th}$ transmitter is $(p_i)^2$. Let us assume the total transmitted energy is $M_t$ (i. e. $\sum_{i=1}^{M_t}(p_i)^2 = M_t$). We assume that there are $K$ targets that are moving in a two dimensional plane. However, without loss of generality, this modeling can be extended to the three dimensional case. The $k^{th}$ target is located at $\mathbf{p}_k = [p_x^k, p_y^k]$ and moves with velocity $\mathbf{v}_k = [\tilde{v}_x^k, \tilde{v}_y^k]$. Now, we model the received signal in four stages as follows:

Stage 1: Under a narrow band assumption on the waveforms, the baseband signal arriving at the $l^{th}$ receiver from the $i^{th}$ transmitter can be expressed as

$$\mathbf{z}_{il}(t) = \sum_{k=1}^{K} \beta_k^{il} p_i x_i(t - \tau_k^{il}) e^{j2\pi(f_k^{il} t - f_c \tau_k^{il})} + \bar{n}_{il}(t) \quad (3)$$

where $\beta_k^{il}$ denotes the attenuation coefficient corresponding to the $k^{th}$ target between the $i^{th}$ transmitter and the $l^{th}$ receiver, $\bar{n}_{il}(t)$ denotes the corresponding received noise, and $f_k^{il}$ and $\tau_k^{il}$ are respectively the corresponding $k^{th}$ target Doppler shift and delay that can be expressed as [19]:

$$f_k^{il} = \frac{f_c}{c} (\mathbf{v}_k \cdot \mathbf{u}_{r_l}^k - \mathbf{v}_k \cdot \mathbf{u}_{t_i}^k) \quad (4)$$

$$\tau_k^{il} = \frac{1}{c} (\|\mathbf{p}_k - \mathbf{t}_i\| + \|\mathbf{p}_k - \mathbf{r}_l\|) \quad (5)$$

where $c$ is the speed of light and $\mathbf{u}_{t_i}^k$ and $\mathbf{u}_{r_l}^k$ denote the unit vector from the $i^{th}$ transmitter to the $k^{th}$ target and unit vector from the $k^{th}$ target to the $l^{th}$ receiver, respectively.

Like [19], we assume that after down converting the received bandpass signal from the radio frequency, it is passed through a bank of $M_t$ matched filters corresponding to $M_t$ transmitters. We assume $\beta_k^{il}$ does not vary within the estimation process duration and the Doppler shift is small (the velocity of targets is much smaller than $c$). Hence, $\beta_k^{il} p_i e^{j2\pi f_k^{il} t}$ can be taken outside of the integral in the matched filter operation. Let us consider $T_s$ as the sampling period time, $T_s \ll T_P$ and $\tau_k^{il} \ll T_P$ for $i = 1, \ldots, M_t, l = 1, \ldots, N_r$, and $k = 1, \ldots, K$. The sampled output of the $i^{th}$ matched filter at the $l^{th}$ receiver in the $m^{th}$ pulse of the estimation process from the $k^{th}$ target can be expressed as

$$z_{il,k}^m(n) = \beta_k^{il} \Psi_{il,k}^m(n) + \bar{n}_{il}(t) * x_i(T_P - t)|_{t=T_P + nT_s} \quad (6)$$

where

$$\Psi_{il,k}^m(n) = p_i e^{j2\pi(f_k^{il}((m-1)T + T_P + nT_s) - f_c \tau_k^{il})} \quad (7)$$

Stage 2: In this stage, at first, we consider that there is only the $k^{th}$ target. Then, we put the output of the matched filters at the $l^{th}$ receiver at a same time in a vector as

$$\mathbf{z}_{l,k}^m(n) = [z_{1l,k}^m(n), \ldots, z_{M_t l,k}^m(n)]^T = \mathbf{\Psi}_{l,k}^m(n) \boldsymbol{\beta}_{l,k} + \mathbf{e}_l^m(n) \quad (8)$$

where

$$e_l^m(n) = [\bar{n}_{1l}^m(t) * x_i(T_P - t) |_{t=T_P+nT_s}, \ldots, \bar{n}_{M_t l}^m(t) \\ * x_i(T_P - t) |_{t=T_P+nT_s}]^T, \quad (9)$$

$$\Psi_{l,k}^m(n) = diag\{\Psi_{1l,k}^m(n), \ldots, \Psi_{M_t l,k}^m(n)\}, \quad (10)$$

$$\boldsymbol{\beta}_{l,k} = [\beta_k^{1l}, \ldots, \beta_k^{M_t l}]^T \quad (11)$$

Next, we put the output vectors of the receivers that are obtained by (8) in vector $\mathbf{z}_k^m(n)$ as

$$\mathbf{z}_k^m(n) = [(\mathbf{z}_{1,k}^m(n))^T, \ldots, (\mathbf{z}_{N_r,k}^m(n))^T]^T \\ = \Psi_k^m(n)\boldsymbol{\beta}_k + \mathbf{e}^m(n) \quad (12)$$

where

$$\mathbf{e}^m(n) = [(\mathbf{e}_1^m(n))^T, \ldots, (\mathbf{e}_{N_r}^m(n))^T]^T, \quad (13)$$

$$\Psi_k^m(n) = diag\{\Psi_{1,k}^m(n), \ldots, \Psi_{N_r,k}^m(n)\}, \quad (14)$$

$$\boldsymbol{\beta}_k = [(\boldsymbol{\beta}_{1,k})^T, \ldots, (\boldsymbol{\beta}_{N_r,k})^T]^T \quad (15)$$

Now, we consider all targets. Putting together the output of all the matched filters of a receiver, and then, putting together the output of all the receivers at a same time in a vector, we have

$$\mathbf{z}^m(n) = \Psi^m(n)\boldsymbol{\beta} + \mathbf{e}^m(n) \quad (16)$$

where

$$\Psi^m(n) = [\Psi_1^m(n), \ldots, \Psi_K^m(n)], \quad (17)$$

$$\boldsymbol{\beta} = [(\boldsymbol{\beta}_1)^T, \ldots, (\boldsymbol{\beta}_K)^T]^T \quad (18)$$

Stage 3: Let us discretize the estimation space and consider it as a four-dimensional space including: the position in the direction $x$, the position in the direction $y$, the velocity in the direction $x$, and the velocity in the direction $y$, which are denoted by $x, y, v_x, v_y$, respectively. The points of this discretized space have the following form.

$$(x_h, y_h, v_x^h, v_y^h) \quad, h = 1, \ldots, L \quad (19)$$

If we define

$$\bar{\boldsymbol{\beta}}_h = \begin{cases} \boldsymbol{\beta}_k, & \text{if the } k^{\text{th}} \text{ target is at } (x_h, y_h, v_x^h, v_y^h) \\ \mathbf{0}_{(M_t N_r) \times 1}, & \text{otherwise} \end{cases} \quad (20)$$

$$\mathbf{s} = [(\bar{\boldsymbol{\beta}}_1)^T, \ldots, (\bar{\boldsymbol{\beta}}_L)^T]^T \quad (21)$$

we can rewrite (15) as

$$\mathbf{z}^m(n) = \widetilde{\Psi}^m(n)\mathbf{s} + \mathbf{e}^m(n) \quad (22)$$

where

$$\widetilde{\Psi}^m(n) = [\widetilde{\Psi}_1^m(n), \ldots, \widetilde{\Psi}_L^m(n)], \quad (23)$$

$$\widetilde{\Psi}_h^m(n) = \Psi_k^m(n)|_{(p_x^k, p_y^k, \tilde{v}_x^k, \tilde{v}_y^k) = (x_h, y_h, v_x^h, v_y^h)} \quad (24)$$

Stage 4: If $N_p$ pulses are used in the estimation process and $N_s$ denotes the number of the achieved samples in each PRI, we have $N_p \times N_s$ samples at the output of each matched filter. Finally, we stack $\{\{\mathbf{z}^m(n)\}_{n=1}^{N_s}\}_{m=1}^{N_p}$ into

$$\mathbf{z}_{(N_p \times N_s \times M_t \times N_r) \times (1)} = [(\mathbf{z}^1(0))^T, \ldots, (\mathbf{z}^1(N_s - 1))^T \\ , \ldots, (\mathbf{z}^{N_p}(0))^T, \ldots, (\mathbf{z}^{N_p}(N_s - 1))^T]^T = \Psi \mathbf{s} + \mathbf{e} \quad (25)$$

where

$$\Psi_{(N_p \times N_s \times M_t \times N_r) \times (L \times M_t \times N_r)} = [(\widetilde{\Psi}^1(0))^T, \ldots, (\widetilde{\Psi}^1(N_s - 1))^T, \ldots, (\widetilde{\Psi}^{N_p}(0))^T, \ldots, (\widetilde{\Psi}^{N_p}(N_s - 1))^T]^T \quad (26)$$

$$\mathbf{e}_{(N_p \times N_s \times M_t \times N_r) \times (1)} = [(\mathbf{e}^1(0))^T, \ldots, (\mathbf{e}^1(N_s - 1))^T \\ , \ldots, (\mathbf{e}^{N_p}(0))^T, \ldots, (\mathbf{e}^{N_p}(N_s - 1))^T]^T \quad (27)$$

Usually, the number of targets is much smaller than the length of $\mathbf{s}$. Hence, $\mathbf{s}$ is a block $K$-sparse vector with the block-length of $d = M_t \times N_r$. We can reconstruct $\mathbf{s}$ from far fewer samples (measurements). These measurements are obtained by multiplying $\boldsymbol{\varphi}_{(M) \times (N_p \times N_s \times M_t \times N_r)}$ with the received signal in each receiver [30]-[32]. Thus, at the fusion center we have

$$\mathbf{y} = \boldsymbol{\varphi}\mathbf{z} = \boldsymbol{\theta}\mathbf{s} + \mathbf{E} \quad (28)$$

where

$$\mathbf{E} = \boldsymbol{\varphi}\mathbf{e} \quad (29)$$

$$\boldsymbol{\theta} = \boldsymbol{\varphi}\Psi \quad (30)$$

$\boldsymbol{\varphi}$ is called the measurement matrix, and $\boldsymbol{\theta}$ is called the sensing matrix. The measurement matrix must be suitable and create small coherence for sensing matrix in classical CS. The coherence of the sensing matrix is the maximum absolute value of the correlation between the sensing matrix columns [13]. It has been shown that zero-mean Gaussian random matrix can be used as a suitable measurement matrix [11].

### III. THE BLOCK RECOVERY ALGORITHMS

There are several block recovery algorithms that can be used for recovering $\mathbf{s}$ in (28). In this paper, the extensions of the Matching Pursuit (MP) and Orthogonal Matching Pursuit (OMP) algorithms [13]-[14] to the block-sparse case are used. These algorithms are named Block Matching Pursuit (BMP) and Block Orthogonal Matching Pursuit (BOMP) [25], respectively.

At first, we divide the sensing matrix into $L$ blocks as

$$\boldsymbol{\theta} = [\underbrace{\boldsymbol{\theta}_1, \ldots, \boldsymbol{\theta}_d}_{\boldsymbol{\theta}[1]}, \underbrace{\boldsymbol{\theta}_{d+1}, \ldots, \boldsymbol{\theta}_{2d}}_{\boldsymbol{\theta}[2]}, \ldots \\ , \underbrace{\boldsymbol{\theta}_{L \times (d-1)+1}, \ldots, \boldsymbol{\theta}_{L \times d}}_{\boldsymbol{\theta}[L]}] \quad (31)$$

where $\boldsymbol{\theta}_l$ is the $l^{\text{th}}$ column of $\boldsymbol{\theta}$.

The BMP algorithm is suggested for the case in which the columns of $\boldsymbol{\theta}[l]$ are orthogonal for each $l$. It should be mentioned that it is not necessary for the columns across different blocks to be orthogonal. BMP algorithm starts by initializing $\mathbf{s}_0 = \mathbf{0}$ and the residual as $\mathbf{r}_0 = \mathbf{y}$. At the $i^{\text{th}}$ stage ($i \geq 1$), according to

$$l_i = \arg \max_l \|\boldsymbol{\theta}^H[l]\mathbf{r}_{i-1}\|_2, \quad (32)$$

the block that is best matched to $\mathbf{r}_{i-1}$ is chosen. Superscript $(.)^H$ denotes the Hermitian of a matrix. After choosing the index $l_i$, $\mathbf{s}[l_i]$ is found by solving

$$\arg\min_{\tilde{\boldsymbol{s}}[l_i]} \|\boldsymbol{y} - \boldsymbol{\theta}[l_i]\tilde{\boldsymbol{s}}[l_i]\|_2, \tag{33}$$

and the residual is directly updated as follows:

$$\boldsymbol{r}_i = \boldsymbol{r}_{i-1} - \boldsymbol{\theta}[l_i]\boldsymbol{\theta}^H[l_i]\boldsymbol{r}_{i-1} \tag{34}$$

Similar to the BMP algorithm, the BOMP algorithm begins by initializing $\boldsymbol{s}_0 = \boldsymbol{0}$ and $\boldsymbol{r}_0 = \boldsymbol{y}$, and according to (32), the best matched block to $\boldsymbol{r}_{i-1}$ is chosen. Next, if $I$ is the set of chosen indices $l_h, 1 \le h \le i$, $\boldsymbol{s}_i$ can be found by solving

$$\arg\min_{\{\tilde{\boldsymbol{s}}_i[l]\}_{l \in I}} \left\| \boldsymbol{y} - \sum_{l \in I} \boldsymbol{\theta}[l]\tilde{\boldsymbol{s}}_i[l] \right\|_2 \tag{35}$$

Finally, we update the residual as

$$\boldsymbol{r}_i = \boldsymbol{y} - \sum_{l \in I} \boldsymbol{\theta}[l]\boldsymbol{s}_i[l] \tag{36}$$

It is clear that BOMP is more complicated than BMP. On the other hand, because of utilizing of all the chosen columns of $\boldsymbol{\theta}$, this method should have better performance.

## IV. THE PROPOSED METHOD FOR ENERGY ALLOCATION

According to [25], we define the block-coherence of two blocks of $\boldsymbol{\theta}$ as:

$$\mu_B^{l,r} = \frac{1}{E_b}\rho(\boldsymbol{\theta}^H[l]\boldsymbol{\theta}[r]) \tag{37}$$

In (37), $\rho(.)$ is the spectral norm which is denoted by $\rho(\boldsymbol{A}) = \sqrt{\lambda_{max}(\boldsymbol{A}^H\boldsymbol{A})}$, where $\lambda_{max}(\boldsymbol{B})$ is the largest eigenvalue of the positive-semidefinite matrix $\boldsymbol{B}$. We have assumed that all the blocks have a same $l_2$-norm, and $\|\boldsymbol{\theta}[l]\|_2 = E_b$ for $l = 1, \dots, L$.

According to section III, for the ideal performance of BMP and BOMP algorithms, $\|\boldsymbol{\theta}^H[r]\boldsymbol{\theta}[l]\boldsymbol{\theta}^H[l]\boldsymbol{\theta}[r]\|_1$ ($\|\boldsymbol{B}\|_1$ is the sum of the absolute values of the elements of $\boldsymbol{B}$) should converge to zero for $l \ne r$. Because if a block of $\boldsymbol{s}$ (for example $\boldsymbol{s}[l]$) is equal to zero, $\boldsymbol{y}$ is not the combination of its corresponding block columns in $\boldsymbol{\theta}$. If $\|\boldsymbol{\theta}^H[r]\boldsymbol{\theta}[l]\boldsymbol{\theta}^H[l]\boldsymbol{\theta}[r]\|_1$ for $l \ne r$ is small enough, we will not choose this block as a candidate of the nonzero-norm blocks. It is shown that for matrix $\boldsymbol{A}_{w \times d}$ with arbitrary $w$, we have [33]:

$$\lambda_{max}(\boldsymbol{A}^H\boldsymbol{A}) \le \max_i \sum_{k=1}^d |(\boldsymbol{A}^H\boldsymbol{A})_{i,k}| \tag{38}$$

Also, we know that

$$\max_i \sum_{k=1}^d |(\boldsymbol{A}^H\boldsymbol{A})_{i,k}| \le \sum_{k=1}^d \sum_{i=1}^d |(\boldsymbol{A}^H\boldsymbol{A})_{i,k}| = \|\boldsymbol{A}^H\boldsymbol{A}\|_1 \tag{39}$$

Thus, we have

$$\lambda_{max}(\boldsymbol{\theta}^H[r]\boldsymbol{\theta}[l]\boldsymbol{\theta}^H[l]\boldsymbol{\theta}[r]) \le \|\boldsymbol{\theta}^H[r]\boldsymbol{\theta}[l]\boldsymbol{\theta}^H[l]\boldsymbol{\theta}[r]\|_1 \tag{40}$$

In order to converge $\|\boldsymbol{\theta}^H[r]\boldsymbol{\theta}[l]\boldsymbol{\theta}^H[l]\boldsymbol{\theta}[r]\|_1$ to zero, it is necessary that $\lambda_{max}(\boldsymbol{\theta}^H[r]\boldsymbol{\theta}[l]\boldsymbol{\theta}^H[l]\boldsymbol{\theta}[r])$ converge to zero. Hence, it is clear that for a good performance of BMP or BOMP, the sum of the block-coherence of the blocks the sensing matrix should converge to zero. Reference [25] has shown that if this measure is small enough, the BMP and BOMP algorithms choose the correct block in each stage. Therefore, we want to allocate optimal energy to the transmitters in order to reduce this value.

Now, we try to achieve an upper bound for the sum of the block-coherence of sensing matrix blocks. According to [34], we can write

$$\mu_B^{l,r} = \frac{1}{E_b}\rho(\boldsymbol{\theta}^H[l]\boldsymbol{\theta}[r]) \le \frac{1}{E_b}\rho(\boldsymbol{\theta}^H[l])\rho(\boldsymbol{\theta}[r]) \tag{41}$$

By using (38) and (39), we have

$$\mu_B^{l,r} \le \frac{1}{E_b}\sqrt{\|\boldsymbol{\theta}^H[l]\boldsymbol{\theta}[l]\|_1}\sqrt{\|\boldsymbol{\theta}^H[r]\boldsymbol{\theta}[r]\|_1} \tag{42}$$

Thus, it is clear that

$$\sum_{l=1}^L \sum_{\substack{r=1 \\ r \ne l}}^L \mu_B^{l,r} \le$$

$$\frac{1}{E_b}\sum_{l=1}^L \sum_{\substack{r=1 \\ r \ne l}}^L \sqrt{\|\boldsymbol{\theta}^H[l]\boldsymbol{\theta}[l]\|_1}\sqrt{\|\boldsymbol{\theta}^H[r]\boldsymbol{\theta}[r]\|_1} \tag{43}$$

Besides, we know

$$(\frac{1}{E_b}\sum_{l=1}^L \sum_{\substack{r=1 \\ r \ne l}}^L \sqrt{\|\boldsymbol{\theta}^H[l]\boldsymbol{\theta}[l]\|_1}\sqrt{\|\boldsymbol{\theta}^H[r]\boldsymbol{\theta}[r]\|_1})^2$$

$$\le \frac{2}{E_b^2}\sum_{l=1}^L \sum_{\substack{r=1 \\ ,r \ne l}}^L \|\boldsymbol{\theta}^H[l]\boldsymbol{\theta}[l]\|_1 \|\boldsymbol{\theta}^H[r]\boldsymbol{\theta}[r]\|_1$$

$$\le 2\left(\frac{1}{E_b}\sum_{l=1}^L \|\boldsymbol{\theta}^H[l]\boldsymbol{\theta}[l]\|_1\right)^2 \tag{44}$$

Therefore, by considering normalized columns for $\bar{\bar{\boldsymbol{\theta}}} = \boldsymbol{\varphi}\bar{\bar{\boldsymbol{\Psi}}}$, the cost function for minimization problem in order to improve the performance of the mentioned block-CS methods can be considered as follows

$$\frac{1}{\sqrt{\sum_{i=1}^{M_t}(p_i)^2}}\sum_{l=1}^L \|\boldsymbol{\theta}^H[l]\boldsymbol{\theta}[l]\|_1 \tag{45}$$

If we define

$$\boldsymbol{H}_1 = diag\{\underbrace{p_1, \dots, p_{M_t}, \dots, p_1, \dots, p_{M_t}}_{N_r}\} \tag{46}$$

$$\bar{\bar{\boldsymbol{\Psi}}} = \boldsymbol{\Psi}|_{(p_1,\dots,p_{M_t})=(1,\dots,1)} \tag{47}$$

the $l^{\text{th}}$ block of $\boldsymbol{\Psi}$ can be written as:

$$\boldsymbol{\Psi}[l] = \bar{\bar{\boldsymbol{\Psi}}}[l]\boldsymbol{H}_1, \tag{48}$$

and the $l^{\text{th}}$ block of $\boldsymbol{\theta}$ can be shown as

$$\boldsymbol{\theta}[l] = \boldsymbol{\varphi}\bar{\bar{\boldsymbol{\Psi}}}[l]\boldsymbol{H}_1, \tag{49}$$

Now, by considering (49), our optimization problem can be expressed as:

$$\min_{p_1,\dots,p_{M_t}} \sum_{l=1}^L \left\|\boldsymbol{H}_1^H\bar{\bar{\boldsymbol{\Psi}}}^H[l]\boldsymbol{\varphi}^H\boldsymbol{\varphi}\bar{\bar{\boldsymbol{\Psi}}}[l]\boldsymbol{H}_1\right\|_1$$

$$\text{s.t.} \sum_{i=1}^{M_t}(p_i)^2 = P_t \tag{50}$$

We should change the problem into a standard form. We do this in three stages as follows.

Stage 1: We know

$$\sum_{l=1}^{L} \|H_1{}^H \bar{\bar{\Psi}}^H[l] \varphi^H \varphi \bar{\bar{\Psi}}[l] H_1\|_1$$

$$= \sum_{l=1}^{L} \|abs(H_1{}^H \bar{\bar{\Psi}}^H[l] \varphi^H \varphi \bar{\bar{\Psi}}[l] H_1)\|_1$$

$$= \|abs(H_1{}^H \sum_{l=1}^{L}(\bar{\bar{\Psi}}^H[l] \varphi^H \varphi \bar{\bar{\Psi}}[l]) H_1)\|_1 \quad (51)$$

$H_1$ is a matrix with real and positive elements. Thus, we can rewrite the cost function of (50) as

$$\left\| H_1{}^H \underbrace{abs(\sum_{l=1}^{L}(\bar{\bar{\Psi}}^H[l] \varphi^H \varphi \bar{\bar{\Psi}}[l])) H_1}_{G} \right\|_1 \quad (52)$$

Elements of $G$ are real and positive. Therefore, (52) is equal to

$$\mathbf{1}_{1 \times d} H_1{}^H \bar{A} H_1 \mathbf{1}_{d \times 1} \quad (53)$$

where $\mathbf{1}_{m \times n}$ is an $m \times n$ matrix with elements that all are equal to 1, and

$$\bar{A} = abs\left(\sum_{l=1}^{L}(\bar{\bar{\Psi}}^H[l] \varphi^H \varphi \bar{\bar{\Psi}}[l])\right) \quad (54)$$

Stage 2: Let us define:

$$\bar{p} = H_1 = [\underbrace{p_1, \ldots, p_{M_t}, \ldots, p_1, \ldots, p_{M_t}}_{N_r}]^T \quad (55)$$

Therefore, the cost function can be expressed as:

$$\bar{p}^H \bar{A} \bar{p} \quad (56)$$

Stage 3: We can define:

$$p = [\tilde{p}_1, \ldots, \tilde{p}_{M_t N_r}]^T, \quad (57)$$

and

$$J_{i,l} = [\bar{J}_i, \underbrace{\mathbf{0}_{1 \times M_t}, \ldots, \mathbf{0}_{1 \times M_t}}_{l-1}, -\bar{J}_i, \underbrace{\mathbf{0}_{1 \times M_t}, \ldots, \mathbf{0}_{1 \times M_t}}_{N_r - l - 1}], \quad (58)$$

$$\bar{J}_i = [\underbrace{0, \ldots, 0}_{i-1}, 1, \underbrace{0, \ldots, 0}_{M_t - i}]. \quad (59)$$

Then, (50) can be rewritten as the following:

$$\min_{p} p^H \bar{A} p$$

$$s.t. \quad J_{i,l} p = 1 \quad \text{for } i = 1, \ldots, M_t$$
$$l = 1, \ldots, N_r - 1,$$
$$\|p\|^2 = P_t N_r \quad (60)$$

Conditions $J_{i,l} p = 1$ (for $i = 1, \ldots, M_t$ and $l = 1, \ldots, N_r - 1$,) are used for equalizing $p$ to $\bar{p}$. In (60), $\bar{A}$ is the sum of some positive semi-definite matrices. Hence, it is also positive semi-definite. Besides, the constraints of this optimization problem are convex except $\|p\|^2 = P_t N_r$. We solve problem (60) without condition $\|p\|^2 = P_t N_r$ and by adding the conditions $\tilde{p}_i \geq p_{min}$ for i = 1, \ldots, M_t N_r$, i.e.,

$$\min_{p} p^H \bar{A} p$$

$$s.t. \quad J_{i,l} p = 1 \quad \text{for } i = 1, \ldots, M_t$$
$$l = 1, \ldots, N_r - 1,$$
$$\tilde{p}_i \geq p_{min} \quad \text{for i} = 1, \ldots, M_t N_r \quad (61)$$

Problem (61) is a convex optimization problem, and for solving it, any proper convex optimization methods can be used. We can also use the CVX software [35],[36]. As it is obvious, the answer of this convex problem ($\hat{p}$) has the lowest cost function among the vectors with the same norm and entries larger than $p_{min}$ that satisfy conditions $J_{i,l} p = 1$. Hence, if we calculate $\alpha$ as $\alpha = \frac{\sqrt{P_t N_r}}{\|\hat{p}\|} \hat{p}$, $\alpha \hat{p}$ has the lowest cost function among the vectors with the norm of $\sqrt{P_t N_r}$ and entries larger than $\alpha p_{min}$ that satisfy conditions $J_{i,l} p = 1$.

## V. THE PROPOSED METHOD FOR DESIGNING THE MEASUREMENT MATRIX

In this section, for designing a proper measurement matrix, we use (45) as the cost function. By defining $F = \varphi^H \varphi$ and allocating uniform energy to the transmitters, the optimization problem becomes:

$$\min_{F} \sum_{l=1}^{L}(\|\Psi^H[l] F \Psi[l]\|_1)$$

$$s.t. \quad \Psi_i{}^H F \Psi_i = 1 \quad for \, i = 1, 2, \ldots, L \times d \quad (62)$$

where $\Psi_i$ is the $i^{th}$ column of the $\Psi$. By defining

$$A_{(Nd)^2 \times Ld} = [vec((\Psi_1 \Psi_1^H)^T), \ldots, vec((\Psi_{Ld} \Psi_{Ld}^H)^T)], \quad (63)$$

we can write the constraints as

$$A^T vec(F) = \mathbf{1}_{(Ld) \times 1}, \quad (64)$$

and the cost function can be written as

$$\sum_{l=1}^{L} \sum_{k=1}^{d} \sum_{\substack{k'=1, \\ k' \neq k}}^{d} abs((vec((\Psi_k[l] \Psi_{k'}^H[l])^T))^T vec(F)) \quad (65)$$

Now, we assume that the elements of $\varphi$ are real and positive. Thus, the elements of $vec(F)$ are also real and positive. Therefore, $abs(vec(F))$ is equal to $vec(F)$, and the optimization problem can be rewritten as

$$\min_{vec(F)} G_t vec(F) \quad s.t. \quad A^T vec(F) = \mathbf{1}_{(Ld) \times 1}$$

$$, G_t = \sum_{l=1}^{L}(\sum_{k=1}^{d} \sum_{\substack{k'=1, \\ k' \neq k}}^{d} abs(((vec\left((\Psi_k[l] \Psi_{k'}^H[l])^T\right))^T)$$
$$(66)$$

In (66), the cost function and the constraint are linear functions. Hence, (67) is a convex optimization problem, and we can solve it with any proper convex optimization methods. Then, $\varphi$ is estimated as

$$\varphi = \Lambda^{0.5} V, \quad (67)$$

where $\Lambda$ is a diagonal matrix with $m$ largest eigenvalues of $F$ as its diagonal elements, and the columns of $V$ are the related eigenvectors of these eigenvalues.

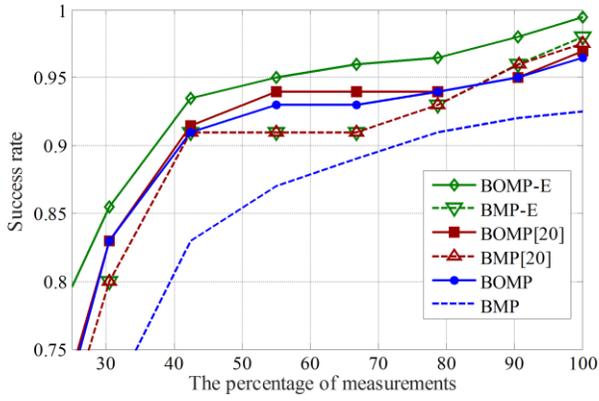

Fig. 1. Success rate versus the percentage of measurements. Here ENR is 10 dB.

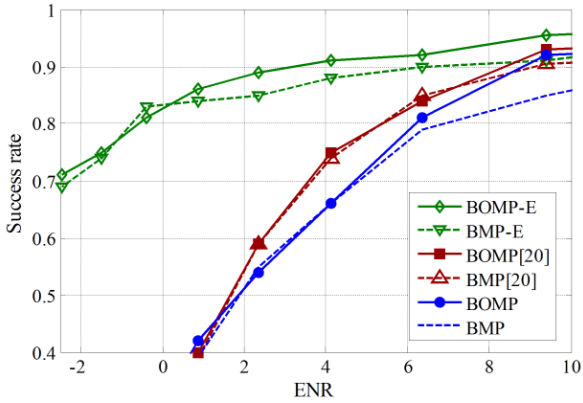

Fig. 2. Success rate versus ENR. Here the percentage of measurement s is 60%.

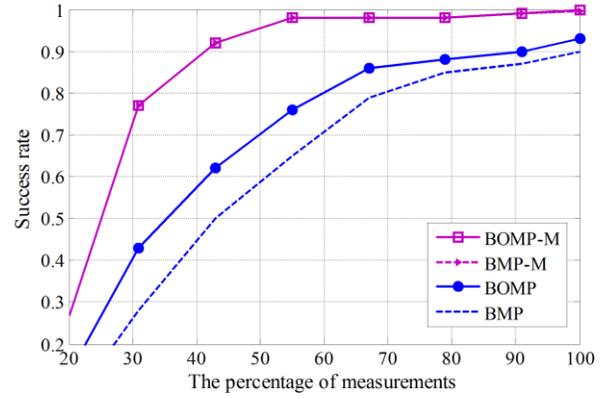

Fig. 3. Success rate versus the percentage of measurements. Here ENR is 10 dB.

$$\boldsymbol{p}_1 = [100,260] \qquad (72)$$

$$\boldsymbol{v}_1 = [120,100] \qquad (73)$$

$$\boldsymbol{p}_2 = [80,280] \qquad (74)$$

$$\boldsymbol{v}_2 = [110,120] \qquad (75)$$

At first, we choose to match the target parameters to the estimation grid in order to focus on the parameter estimation ability of the methods under ideal conditions. The distribution of target attenuation coefficients is complex Gaussian with mean of 0.407 and variance of 0.0907 for both real and imaginary parts, and $\bar{n}_{il}^m(t)$ is zero-mean Gaussian random process.

In this section, we introduce a new measure: total transmitted Energy to the Noise energy Ratio (ENR) defined as

$$\text{ENR} = (M_t \times N_s \times N_p)/(N_s \times N_p \times d \times var_n)$$

$$= \frac{M_t}{d \times var_n} = \frac{1}{N_r \times var_n} \qquad (76)$$

where $var_n$ is the internal noise variance.

Fig. 1 compares the success rate of BOMP, BMP, BOMP-E, BMP-E, BOMP[20], and BMP[20] for different percentages of measurements. BOMP-E and BMP-E are, respectively, related to using BOMP and BMP when the transmitted energy is optimally allocated to the transmitters according to the first proposed method. BOMP and BMP are related to using the mentioned methods when the uniform energy allocation is used, and BOMP[20] and BMP[20] are, respectively, related to the case of using the coherence-based energy allocation that was proposed in [20] and then, exploiting BOMP and BMP algorithms for the recovery. In all the mentioned methods a random Gaussian matrix is used as the measurement matrix. The success rate is the number of the correct estimations of the two targets parameters to the number of total runs. If all the parameters of the targets are estimated correctly, we call it correct estimation. The percentage of measurements is calculated as

## VI. SIMULATION RESULTS

In this section, we demonstrate the performance of block CS-based distributed MIMO radar system using the proposed methods. We consider a distributed MIMO radar system with 2 transmitters and 2 receivers. The transmitters are located at $\boldsymbol{t}_1 = [100,0]$ and $\boldsymbol{t}_2 = [200,0]$, and the receiver locations are $\boldsymbol{r}_1 = [0,200]$ and $\boldsymbol{r}_2 = [0,100]$ (the distance unit is meter). The carrier frequency of the transmitted waveforms is $f_c = 1$ GHz. We choose T= 200 ms, $T_s = 0.2$ ms, $N_s = 10$, and $N_p = 4$. For simplicity, we divide the estimation space into $L = 144$ points (i.e. $(x_h, y_h, v_x^h, v_y^h)$ , $h = 1,…, 144$). The position and velocity of these points are determined as

$$x_h \in \{80,90,100\}, \qquad h = 1,…,144 \qquad (68)$$

$$y_h \in \{260,270,280\}, \qquad h = 1,…,144 \qquad (69)$$

$$v_x^h \in \{100,110,120,130\}, \qquad h = 1,…,144 \qquad (70)$$

$$v_y^h \in \{100,110,120,130\}, \qquad h = 1,…,144 \qquad (77)$$

The velocity unit is m/s. There are 2 targets. The position and velocity of the targets have been chosen as

$$\frac{M}{M_t \times N_r \times N_s \times N_p} \times 100\%$$

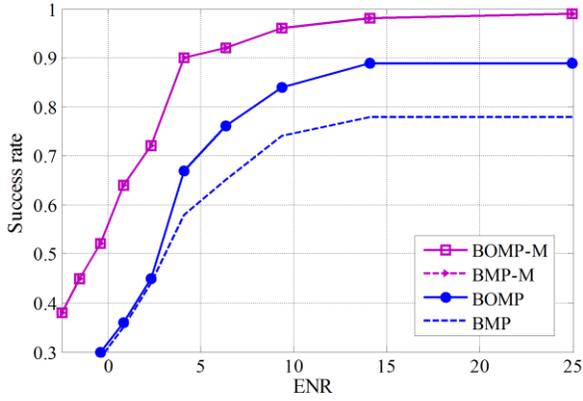

Fig. 4. Success rate versus ENR. Here the percentage of measurements is 60%.

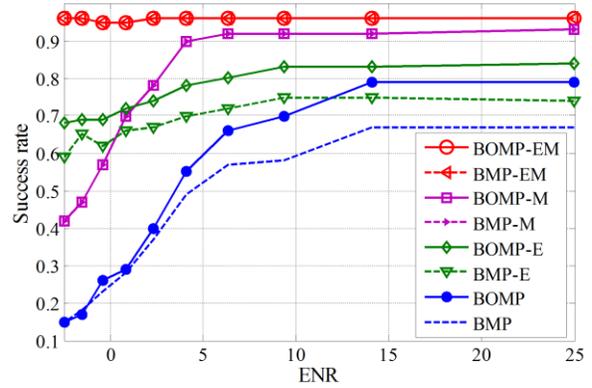

Fig. 6. Success rate versus ENR. Here the percentage of measurements is 50%.

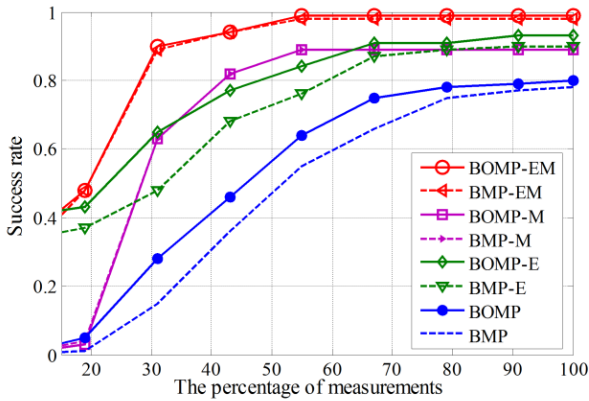

Fig. 5. Success rate versus the percentage of measurements. Here ENR is 5 dB.

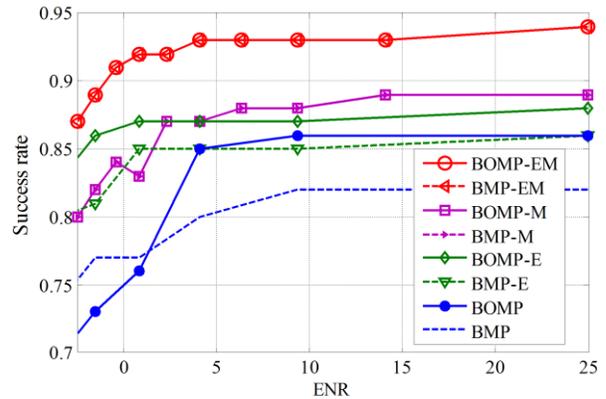

Fig. 7. Success rate versus ENR for off-grid case. Here the percentage of measurements is 50%.

We have used 200 independent runs to generate these results. ENR in this figure is equal to 10 dB, and $p_{min}$ for the proposed energy allocation is 0.1. We can see that using the proposed energy allocation scheme, both BMP and BOMP methods are improved for all different measurement numbers.

Fig. 2 shows the success rate of BOMP, BMP, BOMP-E, BMP-E, BOMP[20], and BMP[20] versus the ENR's. In this figure, the percentage of measurements is equal to 50%. It is clear that the improved methods perform better than the unimproved ones. It can easily be seen from Fig. 2 that using the proposed energy allocation in the block CS-based distributed MIMO radar, more than 80% of the estimations are correct even when the success rates for the coherence-besed and uniform energy allocation methods are less than 0.4.

Now, we consider uniform energy allocation to the transmitters, and for the reduction of complexity $N_p=2$. Fig. 3 compares the success rate of BOMP, BMP, BOMP-M, and BMP-M for different percentages of measurements. BOMP-M and BMP-M are, respectively, related to using BOMP and BMP when the measurement matrix is designed optimally according to the second proposed method. In this figure ENR is 10 dB. We can see that even in the measurement percentage of 35%, using the proposed method for designing the measurement matrix, the success rate is more than 0.8.

Fig. 4 indicates the success rate of BOMP, BMP, BOMP-M, and BMP-M versus ENR. The percentage of measurement is 60% in this figure. As can be seen, by using the optimum measurement matrix, more than 80% of the parameter estimations are correct even in ENR of 3 dB.

Now, we evaluate the performance of a block CS-based distributed MIMO radar using both proposed methods. At first, the measurement matrix is designed by using the second proposed method and the assumption of uniform energy allocation. Then, for the system using this designed measurement matrix, the two transmitter powers are determined according to the first proposed method. BOMP-EM and BMP-EM are the notations that are used for this system when the CS methods are BOMP and BMP, respectively. Fig. 5 plots the success rates of BOMP, BMP, BOMP-E, BMP-E, BOMP-M, BMP-M, BOMP-EM, and BMP-EM for various percentages of measurements with the same assumptions as Figs 3 and 4. The ENR for this Figure is 5 dB. We can see that BOMP-EM and BMP-EM are significantly better than the other methods in all the measurement percentages.

Fig. 6 plots the success rate versus ENR for the mentioned methods. As shown, BOMP-EM and BMP-EM have the best performances among all other methods and their success rates are 0.95 when the success rates of non-improved recovery methods are about 0.15.

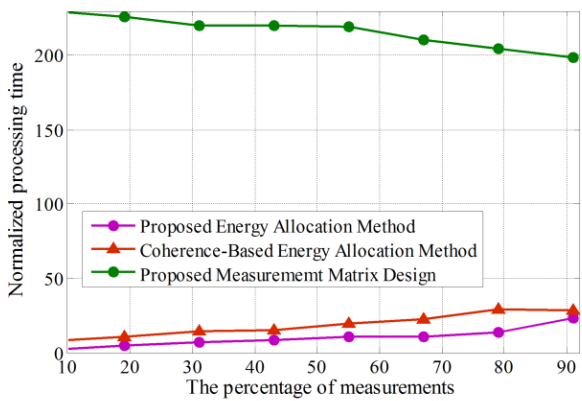

Fig. 8. Normalized processing time versus the percentage of measurements in the ENR of 5.

Now, we assume that the target parameters are out of the estimation grid that is called the off-grid case. We consider the following parameters for our two targets:

$$\boldsymbol{p}_1 = [101, 262] \tag{78}$$

$$\boldsymbol{v}_1 = [123, 104] \tag{79}$$

$$\boldsymbol{p}_2 = [81, 282] \tag{80}$$

$$\boldsymbol{v}_2 = [113, 124] \tag{81}$$

In this case, we consider selecting two nearest points of the estimation grid to the target parameters as the correct estimation. The success rate versus ENR for BOMP, BMP, BOMP-E, BMP-E, BOMP-M, BMP-M, BOMP-EM, and BMP-EM are shown in Fig. 7. As it can be seen, our proposed methods also perform well in this case.

Finally, we evaluate the computation burden of the optimization problems. It should be mentioned that all the mentioned optimizing methods can be done off-line and just once. In Fig. 8, the normalized processing time versus the percentage of measurements for the proposed methods and the energy allocation method of [20] in the ENR of 5 dB are shown. The normalized processing time for each method is obtained by dividing its processing time by the processing time of the BMP method (that is about 1.3 times smaller than that of the BOMP). Figure 8 implies that the energy allocation methods have much lower complexity than the proposed measurement matrix design, and our energy allocation method is better than the coherence based one even in complexity.

## VII. CONCLUSION

We have proposed two methods for the improvement of block CS-based distributed MIMO radar systems. The first method is a transmitted energy allocation scheme, and the second one is an optimal measurement matrix designing method. The proposed methods are based on minimizing an upper bound of the sum of the sensing matrix block-coherences. It has been shown that multiple targets parameters estimation can be improved with the aid of the proposed schemes when the total transmitted energy is constant. Using the first method in the proposed scenario, more than 80% of the estimations are correct even in the ENR of 0 dB when the percentage of measurements is equal to 60%. Exploiting the designed measurement matrix (the second method) in this percentage of measurements, distributed MIMO radars can correctly estimate the multiple targets parameters with the probability of more than 0.9 even when the probability of the correct estimation is less than 0.7 in the non-improved ones. According to the simulation results, combining these proposed methods, the performance improvement is significant. For future works, we will try to improve CS-based distributed MIMO radars in the presence of strong clutter.